# Introducing Coherent-Control Koopman Modeling to Reservoir Scale Porous Media Flow Studies

Dimitrios Voulanas[1,2*], Eduardo Gildin[1]
[1] Harold Vance Department of Petroleum Engineering, Texas A&M University, College Station, TX, USA
[2] Texas A&M Energy Institute, Texas A&M University, College Station, TX, USA
dvoulanas@tamu.edu
*corresponding author

**ABSTRACT**
Accurate and robust surrogate modeling is essential for the real-time control and optimization of large-scale subsurface systems, such as geological $CO_2$ storage and waterflood management. This study investigates the limits of classical Dynamic Mode Decomposition with control (DMDc) and introduces CCKM, as a robust alternative, in enforcing control in pressure and water saturation reservoir dynamics under challenging prediction scenarios. We introduced a control-coherent incremental ($\Delta$) CCKM formulation, in which the field update is driven by actuator changes rather than rather than actuator levels as in the original level formulation and compared them both against DMDc and a Hybrid B-only surrogate that re-uses DMDc's bottom-B (same-step feed-through), showing that only CCKM remains stable and accurate under regime shifts. Two representative cases are considered: (i) an out-of-distribution shut-in and restart case, and (ii) an in-distribution bottomhole pressure (BHP) drawdown. Results show that only CCKM consistently maintains stability and accuracy across both scenarios, achieving sub-bar mean absolute error and sub-percent Frobenius norm percent change error (FPCE) even under regime shifts, while DMDc exhibit large unphysical errors during control transients. The findings demonstrate that strict control-coherence is critical for reliable surrogate modeling, particularly in settings with abrupt changes in control strategy. The proposed framework is broadly applicable to real-time reservoir optimization and can be integrated seamlessly into existing optimization and monitoring workflows, enabling fast and trustworthy decision support in the presence of both expected and unexpected actuation regimes.

Index Terms - Identification for control, subsurface porous media fluid flow, actuator changes, actuator levels, control-coherence

## I. INTRODUCTION

Reservoir management for production optimization and real-time decision support demands fast yet reliable simulation models. High-fidelity reservoir simulations (solving complex multiphase flow PDEs) are often too slow for applications like closed-loop optimization, which may require thousands of forward simulations to find optimal control strategies [1]. Data-driven surrogate models or reduced-order models address this bottleneck by approximating the reservoir dynamics with far less computations, enabling rapid forecasting and control updates. However, capturing the essential nonlinear physics in a surrogate is challenging [1]. Reservoir processes involve multiscale, multiphase interactions, so a surrogate must be accurate over a range of conditions – including scenarios beyond the training data – to be trustworthy at new control sequences.

Over the years, various surrogate modeling approaches have been explored and successfully applied for reservoir simulation. Classical reduced-order models use techniques like Proper Orthogonal Decomposition (POD) or trajectory piecewise linearization to derive low-dimensional models from the governing equations [2], [3]. More recently, data-driven linear surrogates inspired by Koopman operator theory have gained popularity in this domain [4], [1], [5]. Dynamic Mode Decomposition (DMD) and its extension DMD with control [6] offer a way to identify approximate linear dynamics from simulation or field data. The DMDc algorithm, in particular, was introduced to explicitly include the effects of control inputs and disentangle external actuation from the natural reservoir dynamics [6].

In principle, linear surrogates fit historical data well and serve as fast proxies for model predictive control or long-horizon forecasting. However, significant limitations remain. A key concern is generalization. Many data-driven models that match training scenarios closely will accumulate errors over long-term predictions or under unseen inputs [8]. For example, one-step-ahead neural surrogates (e.g. Embed-to-Control networks) tend to drift and degrade when iterated, resulting in poor long-horizon performance. Researchers have found that incorporating multi-step prediction training or physical constraints can mitigate this issue and yield more reliable forecasts [7]. Another critical limitation is the violation of physical causality in naive surrogate identification. A purely regression-based linear model might inadvertently learn a "shortcut" where the control input at time $k$ directly influences the output at time $k$, bypassing the proper state dynamics. Such same-step input-output leakage breaks causality and can occur if the model is overly flexible in how it uses input data. This means the surrogate may simply memorize input/output correlations (through an arbitrary bottom-B block in the learned state-space model) instead of capturing the true reservoir physics. The consequence is that when the surrogate is used for optimization or closed-loop control, it may suggest physically inconsistent actuator trajectories. This undermines trust in the model's predictions for any new control sequence that wasn't in the training data. In summary, current surrogates may be fast, but their tendency to overfit or violate physical principles limits their reliability for both control and forecasting applications in reservoir management.

Control-Coherent Koopman Modeling (CCKM) was recently proposed to eliminate this failure by hard-coding the actuator kinematics and learning only the autonomous response, instead of allowing a regression to freely assign input-to-output influence [8]. By enforcing this structure, the surrogate prevents any direct same-step shortcut from control input to output: control actions can only affect the reservoir state through realistic dynamic pathways that respect time causality. In this letter, we aim to develop a trustworthy, fast surrogate modeling framework for porous media reservoir simulation that enforces exact control over the entire simulation period. We leverage prior knowledge of the actuator dynamics (kinematics) – how control actions (e.g. BHP changes, injection rates) propagate through immediate fluid responses – and explicitly incorporate this known behavior into the surrogate instead of allowing a regression to freely assign input-to-output influence. The remaining part of the surrogate, representing the reservoir's autonomous (actuator/well – reservoir coupling for Δ formulation) response, is then learned from data (for example, learning the evolution of pressure and saturation states driven by reservoir physics).

This letter is structured as follows: Section II develops dynamical-system identification: defines the reservoir–actuator augmented state, states the identification objective, and presents both standard DMDc and the Control-Coherent Koopman models (level and proposed Δ formulations) with the causality constraint (zero bottom–B). Section III specifies the primary-variable estimation study: MRST domain, counterexamples (shut-in/restart; BHP drawdown), train/test split, and error metrics. Section IV reports results and discusses accuracy and actuator consistency across scenarios. Section V concludes and outlines applications to multi-control/multi-phase and real-time optimization.

## II. DYNAMICAL SYSTEM IDENTIFICATION

The dynamics of a high-dimensional, discrete system and an actuation sub-system are described by:
$$x_{k+1} = f_x(x_k, u_k) \quad (1)$$
$$p_{k+1} = f_p(x_k, u_k) \quad (2)$$
where $x_k \in \mathbb{R}^{N_c}$ is the state (e.g., the pressure or saturation field), $u_k \in \mathbb{R}^m$ the control input, $p_k \in \mathbb{R}^m$ is the actuator state (e.g., cumulative injected volumes or well BHPs), $f(\cdot)$ is the nonlinear, in general, operator that propagates the state $x_k$ by one time step.

The goal of this letter is to identify a linear state space model of Eq. (1) that is suitable for exact control enforcement, using experimental or numerical data only (i.e., without the need to know the underlying operators $f(\cdot)$ and their exact relationship with given controls.

Given a set of training states for both well or actuator ($p_k$) and reservoir ($x_k$) at time $k$, arranged as:
$$z_k = [p_k, x_k]^\top \in \mathbb{R}^{m+N_c} \quad (3)$$
$$z_{k+1} = [p_{k+1}, x_{k+1}]^\top \in \mathbb{R}^{m+N_c} \quad (4)$$
Identification on the augmented states yields the following one-step affine propagator:
$$z_{k+1} = Az_k + Bu_k + c \quad (5)$$

### A. STANDARD DMD WITH CONTROL IDENTIFICATION

DMDc and related techniques identify all blocks of $A$ and $B$ by regression:
$$z_{k+1} = \begin{bmatrix} A_{pp} & A_{px} \\ A_{xp} & A_{xx} \end{bmatrix} z_k + \begin{bmatrix} B_p \\ B_x \end{bmatrix} u_k \quad (6)$$
where all blocks are fit jointly from data. This allows $u_k$ to immediately affect $z_{k+1}$ through the lower block of $B$ ("bottom–B"), bypassing the actuator state. Such a same-step input-to-field shortcut breaks causality and can lead to actuator-inconsistent field predictions, particularly under controls not seen in training. Normally, DMDc only requires $(x_k, x_{k+1}, u_k)$ to be given as training data. Here, however, we choose the augmented state form to create a fair comparison setting.

### B. ACTUATOR KINEMATICS IN CONTROL-COHERENT KOOPMAN RESERVOIR MODELING

Control-Coherent Koopman Modeling (CCKM) [8] enforces physical causality by separating known actuator dynamics from unknown reservoir response, whose primary assumption is that the overall system dynamics can be cleanly partitioned into a known actuator sub-system (with explicitly specified kinematics) and an unknown autonomous sub-system (to be learned from data), as mentioned earlier. The actuator/well state $p_k$ represents the physical variables that are directly set by the control - such as the cumulative injected volume (in rate-controlled settings) or well bottomhole pressure (in BHP-controlled wells).

To obtain a one-step equivalent formulation where both $p_{k+1}$ and $x_{k+1}$ are computed simultaneously, we proceed as follows:

$$A_{pp} = \begin{cases} I, & \text{(rate mode)} \\ (1-\lambda)I, & \text{(BHP mode)} \end{cases} \quad (7)$$

$$B_p = \begin{cases} \Delta t\, I, & \text{(rate mode)} \\ \lambda I, & \text{(BHP mode)} \end{cases} \quad (8)$$

with $I$ the $m \times m$ identity. Under rate control, the cumulative injection follows $p_{k+1} = p_k + \Delta t\, u_k$, where $u_k$ is the prescribed rate at step $k$ and $\Delta t$ is the timestep. Similarly, for BHP control with a first-order well response of gain $\lambda$ (leaky integrator), we have $p_{k+1} = p_k + \lambda(u_k - p_k)$. The standard Peaceman well model [9], which connects well and reservoir, is built on a steady near-well reference solution and it neglects wellbore dynamics so that the commanded BHP equals the pressure right at the well/reservoir border (sandface). Under that assumption, the control state must pass through unfiltered, so using the leaky integrator with $\lambda = 1$ is the consistent choice.

### C. CCK IDENTIFICATION: LEVEL UPDATE

For variable like pressure evolving under BHP the level formulation is used. For simultaneous update (i.e., to make both $p_{k+1}$ and $x_{k+1}$ depend on $z_k$ and $u_k$), one may algebraically note
$$p_k = A_{pp}^{-1}(p_{k+1} - B_p u_k) \quad (9)$$
but to preserve causality (i.e., avoid using $p_{k+1}$ in the update of $x_{k+1}$), we equivalently write the full block update as:

$$z_{k+1} = \begin{bmatrix} A_{pp} & 0_{m \times N_c} \\ A_{xp}^{(lvl)} & A_{xx}^{(lvl)} \end{bmatrix} z_k + \begin{bmatrix} B_p \\ 0_{N_c \times m} \end{bmatrix} u_k \quad (10)$$

with the actuator and field state updates, respectively:
$$p_{k+1} = A_{pp} p_k + B_p u_k \quad (11)$$
$$x_{k+1} = A_{xp}^{(lvl)} p_k + A_{xx}^{(lvl)} x_k \quad (12)$$

The "bottom–B" block is fixed to zero (see second term in Eq. (10)), as stated earlier, and the actuator-to-field coupling (i.e., $A_{xp}$) are learned from data. This level form ensures that control input can affect the field only via the physically consistent actuator pathway, not through a same-step shortcut.

### D. CCK IDENTIFICATION: INCREMENTAL (DELTA) UPDATE

For variables (e.g., saturation) where increments are physically meaningful or the actuator/well state evolves incrementally, we introduce an incremental (delta) formulation. The field evolution is written in terms of increments:

$$x_{k+1} - x_k = A_{xp}^{(\Delta)}(p_{k+1} - p_k) + A_{xx}^{(\Delta)} x_k + b_x \quad (13)$$

or equivalently,
$$x_{k+1} = x_k + A_{xp}^{(\Delta)} \Delta p_k + A_{xx}^{(\Delta)} x_k + b_x \quad (14)$$

where $\Delta p_k = p_{k+1} - p_k$ is computed from the known actuator kinematics (rate/BHP), and $b_x$ is a bias term to absorb steady-state offsets.

The full augmented state update is then:

$$\begin{bmatrix} p_{k+1} \\ x_{k+1} \end{bmatrix} = \begin{bmatrix} A_{pp} & 0_{m \times N_c} \\ A_{xp}^{(\Delta)}(A_{pp} - I) & I + A_{xx}^{(\Delta)} \end{bmatrix} \begin{bmatrix} p_k \\ x_k \end{bmatrix} \\ + \begin{bmatrix} B_p \\ A_{xp}^{(\Delta)} B_p \end{bmatrix} u_k + \begin{bmatrix} 0 \\ b_x \end{bmatrix} \quad (15)$$

with rate mode: $\Delta p_k = \Delta t\, u_k$ and BHP mode: $\Delta p_k = \lambda(u_k - p_k)$.

This parameterization is advantageous in capturing rapid transitions or when the system response is primarily a function of actuator increments rather than levels.

**Remark (Rate mode)**: With constant rate control, $\Delta p_k = \Delta t\, u_k$ is nonzero each step; the $\Delta$ formulation correctly treats a steady injection/production schedule as a sustained incremental drive.

**Remark (Constant BHP)**: The field state continues to evolve through the learned coupling and autonomous dynamics, so a drawdown ($p_k$ held below reservoir pressure) yields saturation change via the level form terms.

### III. PRIMARY VARIABLE ESTIMATION

To evaluate the practical performance of control-coherent surrogates in reservoir settings, we construct two test cases representing simple field scenarios. These settings are used both to test primary variable (pressure and saturation) estimation, and to design counterexamples that expose the failure of unconstrained DMDc, contrasting them with the actuator-consistent predictions of CCKM.

#### A. DOMAIN AND PROBLEM SETUP

All simulations are performed using a standard two-dimensional Cartesian reservoir model, discretized on a square cartesian grid of size $n_x \times n_x$, with physical dimensions $2000 \times 2000 \times 20\, m$ and uniform properties (100 md permeability, 0.25 porosity, linear relative permeability curves). The underlying simulator is the MRST ad-blackoil module [10], which provides fully coupled multiphase flow and well control functionality. For each scenario, only a single well is placed at the grid center (odd $n_x$), and appropriate boundary conditions (no-flow) are used to isolate the well response. Fig. 1 depicts the reservoir shape and outline along with the well location.

Multiphase flow equations are based on the principles of mass conservation and Darcy's law. For $N$ fluids with no mass transfer between phases (immiscible) that each consists of a single component, the conservation equation per phase becomes:

$$\frac{\partial}{\partial t}(\varphi \rho_\alpha S_\alpha) + \nabla \cdot (\rho_\alpha \vec{u}_\alpha) = \rho_\alpha q_\alpha \quad (16)$$

where $\varphi$ is porosity (This is a measure of the fraction of the total volume of the porous medium that is occupied by pores), $\rho_\alpha$ is the density of phase $\alpha$, $S_\alpha$ is the saturation of phase $\alpha$ (This is the fraction of the pore volume that is occupied by phase $\alpha$, and have value between 0 and 1 inclusive. Note that the summation of all phase saturations have to equal 1), $\frac{\partial}{\partial t}(\varphi \rho_\alpha S_\alpha)$ represents the rate of change of mass of phase $\alpha$ per unit volume with respect to time, $\nabla \cdot$ represents the net rate at which the mass flux of phase $\alpha$ is flowing out of a given volume, $\vec{u}_\alpha$ is the Darcy velocity of phase $\alpha$, $\nabla \cdot (\rho_\alpha \vec{u}_\alpha)$ represents the net mass flux of phase $\alpha$ into or out of a control volume, and $q_\alpha$ is the volumetric source term for phase $\alpha$, which can include sources (injection) or sinks (production).

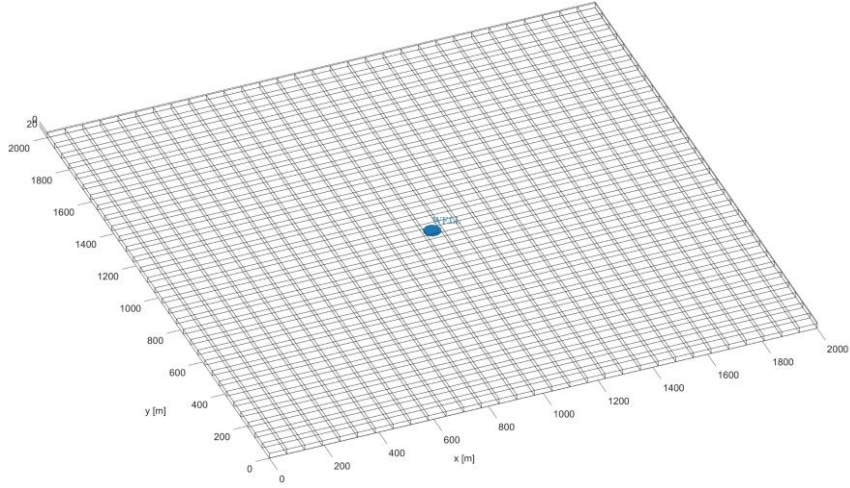

Fig. 1 – Reservoir Shape and Outline and Well Location

In addition, the primary relationship used to form a closed model is Darcy's law, which can be extended to multiphase flow by using the concept of relative permeabilities.

$$\vec{u}_\alpha = -\frac{K \kappa_{r\alpha}}{\mu_\alpha}(\nabla p_\alpha - g\rho_\alpha \nabla z) \tag{17}$$

where $K$ is the absolute permeability of the porous medium, $\kappa_{r\alpha}$ is the relative permeability of phase $\alpha$, $\mu_\alpha$ is the viscosity of phase $\alpha$, $\nabla p_\alpha$ is the pressure gradient of phase $\alpha$, $g$ is the gravity acceleration, and $\nabla z$ is the gradient in the vertical direction.
This extension of Darcy's law to multiphase flow is often attributed to [11], [12].

### B. COUNTEREXAMPLE DESIGN

To emphasize the importance of actuator-consistent modeling, we design control tests that stress the models' ability to handle out-of-distribution actuator trajectories. For a rate-controlled injector, two distinct input sequences are used for testing: (1) a single central injector is operated in rate control. The training period consists of low rate, steady injection. In the test period, the well is first shut in (0 rate) for an extended interval, followed by a high-rate injection phase. (2) a single central producer operates in BHP control. The training period consists of 90 bars BHP pressure drop from 200 bars of initial reservoir pressure. During the test period, the BHP setpoint is sharply reduced again by 90 bars and held at the new low value (Fig. 2 – 1B).

### C. EXPERIMENTAL PROTOCOL

For each case, the data is split into training (identification) and testing (generalization) periods, with the well control protocol (rate or BHP) and simulation time steps ($\Delta t$) matched across all surrogate models. The full field states (pressure and saturation) are recorded at every step, and both DMDc and CCKM models are identified using the training data only. We evaluate all models on the unseen control sequences, focusing on: (1) primary variable accuracy: Measured as mean absolute error (pressure/saturation) and normalized Frobenius norm (pressure only), (2) control channel consistency, which is the discrepancy between the predicted actuator trajectory (cumulative injected volume or BHP) and the schedule set by the physical control protocol, quantifying whether the model respects the known actuator dynamics. Through these experiments, we demonstrate that DMDc, despite fitting training data, can produce actuator-inconsistent and non-physical field trajectories under challenging test protocols. In contrast, the CCKM surrogate, by design, remains actuator-consistent and provides robust primary variable estimation across all scenarios.

## IV. RESULTS

This section reports the performance of CCKM on the two representative scenarios described earlier: (i) an out-of-distribution shut-in/restart experiment (Case A), and (ii) a standard BHP drawdown under in-distribution conditions (Case B). Figs. 2 and 3 visualize the dynamic responses of the primary variables (pressure and water saturation) and the control/actuator channels, while Table 1 summarizes the test metrics.

### A. Training Performance

DMDc and CCKM (full order) achieved very low errors on the training data, with MAE less than $10^{-5}$ and FPCE % under 0.001%. This confirms that all models represent the training trajectories accurately in a least-squares sense. No $A_{xx}$ drift or instability was ever observed in any of the models.

### B. Discussion

In Case A (shut-in and restart), DMDc continued the training-period rise during the shut-in/restart and performed poorly (FPCE ~93.0%), consistent with its free same-step input path and lack of actuator-law enforcement while Hybrid B-only yielded 2361% FPCE. Here, Hybrid-B fails because the net same-step gain $g = B_{\text{DMDc}} - A_{xp}^{(\text{CCKM - }\Delta)}B_p$ is large ($\| g \|_F \approx 4.66$) and is directly multiplied by the 100× input spike ($q_{test} = 100 q_{train}$) (Fig. 2 – 1A). In contrast, CCKM ($\Delta$) followed the simulated testing period closely with 0.07% FPCE. The control channel is honored by both CCKM ($\Delta$) and Hybrid B-only pressure models in contrast to DMDc which has a 94.6% FPCE (Fig. 3 – 1A). Regarding Case A water saturation, Hybrid B-only overshoots above 1 saturation catastrophically after the shut-in period while DMDc undershoots the shut-in period and breaks down during restart, yielding 139 and 4.99 MAE, respectively (Fig. 2 – 2A). Quantitatively, DMDc relies on $\| B_{\text{bottom}} \|_F \approx 0.283$ while the actuator-mediated path is negligible ($\| A_{\psi p} B_p \|_F \approx 3.97 \times 10^{-4}$), making the prediction overly sensitive to input spikes despite reasonable actuator consistency (0.283%). Hybrid-B's failure is explained by the residual same-step feedthrough; numerically $\| g \|_F \approx 0.282$, orders of magnitude larger than the coherent actuator path ($\| A_{xp}^{(\text{CCKM})} B_p \|_F \approx 8.6 \times 10^{-4}$).

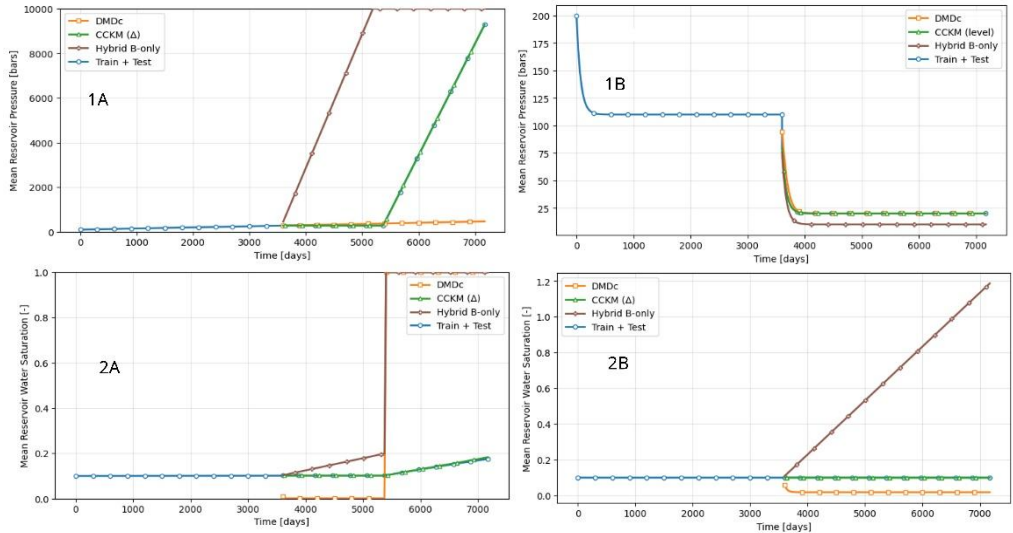

Fig. 2 – Train and Test (Simulated and Surrogate Model Predictions) Mean Reservoir Primary Variables. 1/2 A and 1/2 B represent the Mean Reservoir Pressure and Water Saturation of Case 1 and Case 2, respectively.

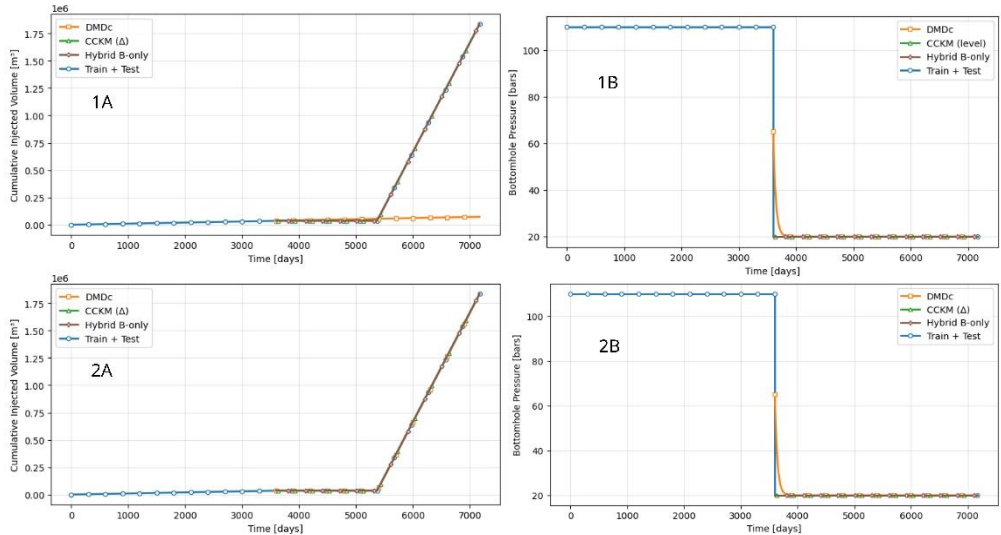

Fig. 3 – Train and Test (Simulated and Surrogate Model Predictions) Control Channels. 1/2A and 1/2B represent the Cumulative Injected Volume and Bottomhole Pressure of Case 1 and Case 2, respectively.

In contrast, CCKM ($\Delta$) remains stable and accurately tracks test water saturation with 0.07 MAE (Fig. 2 – 2A. The relatively high FPCE is attributed to sparsity of the water saturation variable, in which the tiny denominator makes the metric seem large). All water saturation models capture the control channel (Fig. 3 – 2A). Similarly, for Case B (BHP drawdown), CCKM (level/$\Delta$) again provides robust predictions, with pressure MAE less than 0.01 bars and less than 0.1 FPCE. Hybrid B-only and DMDc, however, show significant pressure test errors, FPCE over 10% and 40%, respectively (Fig. 2 – 1B).

DMDc pressure and water saturation control channel over predicts by 23.73% but eventually catches up with CCKM (level) and Hybrid B-only (Fig. 3 – 1B/2B). DMDc water saturation shows relative error of 0.08 MAE while Hybrid B-only shows significantly higher error of 0.55 MAE (Fig. 2 – 2B). CCKM (Δ) error is negligible. These results highlight a key implication: CCKM (level/Δ) is uniquely robust to aggressive and out-of-distribution actuator signals because it preserves actuator-state coherence. By contrast, DMDc and Hybrid B-only, which inherit DMDc's "bottom-B" feedthrough, are prone to error amplification whenever test controls depart from those seen in training.

Table 1 - Prediction Performance for Pressure and Water Saturation

| Case 1 - P/$S_W$ | MAE [bars/-] | FPCE [%] |
|---|---|---|
| DMDc | 2240 / 4.99 | 93 / 4895 |
| CCKM (Δ) | 0.83 / 0.03 | 0.07 / 396 |
| Hybrid B-only | 5.4e+04 /139 | 2361 / 159073 |
| Case 2 - P/$S_W$ | MAE [bars/-] | FPCE [%] |
| DMDc | 0.75 / 0.08 | 13.47 / 81.21 |
| CCKM (level/Δ) | 0.007 / 0.0001 | 0.03 / 0.13 |
| Hybrid B-only | 9.84 / 0.55 | 43.78 / 634 |

## V. CONCLUSIONS

This study demonstrates that strict control-coherent surrogate modeling, as implemented in the CCKM framework and its proposed extension, is essential for reliable dynamic forecasting in subsurface reservoir systems under both typical and extreme actuation scenarios. CCKM surrogate models consistently outperform classical DMDc and the hybrid B-only surrogates, maintaining physical actuator consistency and low prediction errors, even when the control regime is significantly different from the training data. These results highlight the risks of unstructured reduced-order surrogates and the necessity of enforcing actuator-physics constraints. The approach is readily extensible to multi-control, multi-phase settings, and is directly applicable to real-time optimization and subsurface monitoring workflows for optimal subsurface management.

**Declaration of Generative AI and AI-assisted technologies in the writing process**

During the preparation of this work the authors used ChatGPT to explore possible improvements in the authors' text. After using this tool, the authors reviewed and edited the content as needed and take full responsibility for the content of the publication.

## REFERENCES

[1] A. Bao, E. Gildin, A. Narasingam, and J. S. Kwon, "Data-Driven Model Reduction for Coupled Flow and Geomechanics Based on DMD Methods," *Fluids*, vol. 4, no. 3, p. 138, July 2019, doi: 10.3390/fluids4030138.

[2] J. He and L. J. Durlofsky, "Reduced-order modeling for compositional simulation by use of trajectory piecewise linearization," *SPE Journal*, vol. 19, no. 5, pp. 858–872, 2014, doi: 10.2118/163634-pa.

[3] Y. Yang, M. Ghasemi, E. Gildin, Y. Efendiev, and V. Calo, "Fast multiscale reservoir simulations with POD-DEIM model reduction," *SPE Journal*, vol. 21, no. 6, pp. 2141–2154, 2016, doi: 10.2118/173271-PA.

[4] H. Zalavadia, S. Sankaran, M. Kara, W. Sun, and E. Gildin, "A Hybrid Modeling Approach to Production Control Optimization Using Dynamic Mode Decomposition," in *SPE Annual Technical Conference and Exhibition*, Calgary, Alberta, Canada: SPE, Sept. 2019, p. D022S090R002. doi: 10.2118/196124-MS.

[5] D. Voulanas and E. Gildin, "Dynamic mode decomposition accelerated forecast and optimization of geological CO2 storage in deep saline aquifers," *Computers & Chemical Engineering*, vol. 204, p. 109377, Jan. 2026, doi: 10.1016/j.compchemeng.2025.109377.

[6] J. L. Proctor, S. L. Brunton, and J. N. Kutz, "Dynamic Mode Decomposition with Control," *SIAM Journal on Applied Dynamical Systems*, vol. 15, no. 1, pp. 142–161, Jan. 2016, doi: 10.1137/15M1013857.

[7] J. Chen, E. Gildin, and J. Killough, "Multi-Step Embed to Control: A Novel Deep Learning-based Approach for Surrogate Modelling in Reservoir Simulation," Oct. 12, 2024, *arXiv*: arXiv:2409.09920. doi: 10.48550/arXiv.2409.09920.

[8] H. H. Asada and J. A. Solano-Castellanos, "Control-Coherent Koopman Modeling: A Physical Modeling Approach," in *2024 IEEE 63rd Conference on Decision and Control (CDC)*, Milan, Italy: IEEE, Dec. 2024, pp. 7314–7319. doi: 10.1109/CDC56724.2024.10886771.

[9] D. W. Peaceman, "Interpretation of Well-Block Pressures in Numerical Reservoir Simulation(includes associated paper 6988 )," *Society of Petroleum Engineers Journal*, vol. 18, no. 03, pp. 183–194, June 1978, doi: 10.2118/6893-PA.

[10] K.-A. Lie, "Reservoir Modelling Using MATLAB-The MATLAB Reservoir Simulation Toolbox (MRST)," no. November, pp. 17–18, 2020.

[11] M. Muskat and M. W. Meres, "The Flow of Heterogeneous Fluids Through Porous Media," *Physics*, vol. 7, no. 9, pp. 346–363, Sept. 1936, doi: 10.1063/1.1745403.

[12] M. Muskat, R. D. Wyckoff, H. G. Botset, and M. W. Meres, "Flow of Gas-liquid Mixtures through Sands," *Transactions of the AIME*, vol. 123, no. 01, pp. 69–96, Dec. 1937, doi: 10.2118/937069-G.